# NB-IoT Uplink Receiver Design and Performance Study

Arvind Chakrapani, Qualcomm Flarion Technologies, New Jersey, USA.
Email: achakrap@qti.qualcomm.com

*Abstract*— **LTE Narrowband Internet of Things (NB-IoT) is a 3GPP[1] defined cellular technology that is designed to enable connectivity to many low-cost and low power/throughput IoT devices running delay-tolerant applications. NB-IoT can coexist within LTE spectrum either in a standalone mode, in-band with LTE or in guard-band of LTE. With NB-IoT designed to operate in very low signal to noise power ratios, the uplink receiver design presents several challenges. In this paper the design and performance of a NB-IoT uplink receiver is studied in detail. First, receiver design for NB-IoT uplink channels, with corresponding mathematical analysis, is presented. Specifically, it is shown how the time/frequency structure of signals can be exploited to enhance the receiver performance. Second, the performance of each channel is characterized with both link level simulations and implementation on a commercially deployed Qualcomm® FSM™ platform[2] [1]. Comparisons against the 3GPP defined Radio Performance and Protocol aspect requirements[3] are also provided. Finally, implementation details are addressed and discussions on proposed enhancements to NB-IoT in 3GPP Release 15 are provided. It is shown how the proposed receiver algorithms can be adopted to Release 15 enhancements with minor or no modifications. The work in this paper is of significance to system designers looking to implement efficient NB-IoT uplink receiver to coexist with legacy LTE systems.**

## 1. Introduction

Internet of Things (IoT) will connect numerous heterogeneous devices to the internet; enabling data exchange, processing, monitoring, control, etc. resulting in a myriad of use cases. Cellular technologies have been developed or being enhanced to play an indispensable role in the IoT world by catering to categories of devices with different capabilities in power, cost and complexity. Narrowband IoT (NB-IoT), developed (and being enhanced) by 3GPP, promises to provide improved coverage for massive number of low-throughput low-cost devices having low power consumption requirements [2]-[5] with relaxed latency constraints. Prospective applications include utility metering, environment monitoring, asset tracking, municipal light and waste management, etc. The NB-IoT downlink (DL) and uplink (UL) channels are restricted to 180kHz system bandwidth [7], occupying only one physical resource block (PRB) of LTE and supported only for frequency-division multiplexing (FDD) mode in 3GPP releases 13 and 14. NB-IoT can operate in in-band or guard-band of LTE or in a standalone manner. The NB-IoT DL uses the same numerology as the 3GPP Long Term Evolution (LTE) and is based on the LTE design of orthogonal frequency division multiplexing (OFDM). The NB-IoT UL is based on a single-carrier frequency-division multiple access (SC-FDMA) similar to LTE, with some important changes (see [3]). The NB-IoT uplink has three physical channels defined with a completely redesigned random access (NPRACH) along with data (NPUSCH Format 1) and control (NPUSCH Format 2) channels. The NB-IoT uplink channels can be configured to use different subcarrier spacings[4,] single or multiple tones and repetitions for providing operating flexibility. With NB-IoT designed to cater for a maximum coupling loss (i.e. enhanced coverage) of up to 164dB in 3GPP release 14 [2], the operating signal-to-noise ratio (SNR) regimes are such that signal power is well below noise power requiring receiver to boost the received SNR through time/frequency combining. In this paper, the focus is on the design and performance of an NB-IoT uplink receiver.

Although, there have been several contributions in 3GPP that reported the receiver performance of the NB-IoT uplink channels, receiver design details have not been provided (see [2] and the references therein). As far as we are aware, the receiver design details for NB-IoT uplink shared data and control channels (i.e., NPUSCH Formats 1 and 2 respectively) have not been investigated in the current literature. However, recently there has been a few studies of the NB-IoT random access channel (NPRACH) receiver [4]-[7]. An NPRACH receiver was proposed in [4], where a joint estimation of preamble detection and timing delay estimation was proposed using a two-dimensional FFT. However, no discussion is provided on how to derive the threshold for correct detection and implementation issues are not addressed. Moreover, the complexity of doing a 2-dimensional FFT in a practical system is prohibitive. A joint optimization technique was investigated in [6] to configure the NPRACH parameters in order to reduce the collision probability under a target delay constraint.

However, the complexity of the proposed scheme may not be feasible for practical implementation, particularly when NB-IoT is required to co-exist with LTE and multiple users of both LTE and NB-IoT have to be served. In [8], a low-complexity NPRACH receiver was proposed. The proposed method incurs a $4 \times R$ delay in detection and timing estimation, where $R$ is the configured repetitions of the preambles. The delay in preamble detection could become very large when the number of configured repetitions is large. Moreover, the storage requirements scale with repetitions configured (see [8] for more details). In a real-time system with processing and memory constraints, a simpler receiver algorithm is desirable. In this work efficient receiver algorithms for all three NB-IoT uplink channels are proposed with the following contributions. First, a detailed mathematical analysis is provided with receiver design guidelines for each NB-IoT uplink channel. It is shown how the similarity between NPUSCH single tone data channel and NPUSCH control channel can be exploited to use common estimation modules to reduce memory and implementation efforts. Second, discussions are provided on implementation issues to be considered in a practical system. Third, link level simulation results are presented along with performance study on a Qualcomm FSM Chipset based LTE small cell base-station [1]. The rest of the document is arranged as follows. In section 2, the NPRACH channel is introduced and a receiver solution proposed. In section 3, the uplink shared channel for data i.e., NPUSCH Format 1 receiver is analyzed followed by an analysis and design of the NPUSCH Format 2 receiver in section 4, where most of the estimation modules of NPUSCH Format 1 are reused. In section 5, the performance of each of the NB-IoT channels is provided using detailed link-level simulations followed by discussions on implementation complexity and practical issues. Comparison of the simulation results with RAN4 recommended minimum performance requirements is provided along with NB-IoT uplink receiver performance on a Qualcomm FSM Chipset based LTE small cell base-station. Finally, in section 6, discussions on NB-IoT enhancements being proposed in 3GPP release 15 are provided and how the proposed receiver can be adopted for the enhancements are shown.

---

[1] NB-IoT was finalized as part of 3rd Generation Partnership Project (3GPP) release 14 specification[2].

[2] Qualcomm FSM is a product of Qualcomm Technologies, Inc. and/or its subsidiaries [1].

[3] Radio Performance and Protocol aspect requirements are specified by the 3GPP Radio Access Network working group 4, also known as RAN4.

[4] For NPRACH, only 3.75kHz subcarrier spacing is allowed in 3GPP release 14 and can be configured to use either 1.25kHz or 3.75kHz in release 15.

## 2. NPRACH RECEIVER DESIGN

Transmitting a random-access preamble is the first step of random-access procedure that enables a user equipment (UE) to establish a connection with the network. Apart from accurate preamble detection, estimating uplink timing is another main objective of NPRACH receiver. The acquired uplink timing is used to command the UE to perform timing advance to achieve uplink orthogonalization in OFDMA/SCFDMA systems [3]. The NB-IoT physical random-access channel (NPRACH) refers to the time frequency resource on which random access preambles are transmitted. The required NPRACH parameters configured by higher layers [13] are illustrated in Figure 2-1 and defined as,

1. $N_{period}^{NPRACH}$ (3 bits, *nprach-Periodicity*) ∈ {40, 80, 160, 240, 320, 640, 1280, 2560} in milliseconds corresponds to the NPRACH period over which UEs can do random access.
2. $N_{rep}^{NPRACH}$ (3 bits, *nprach-NumRepetitions*) ∈ {1, 2, 4, 8, 16, 32, 64, 128}, corresponds to the NPRACH preamble repetitions.
3. $N_{scoffset}^{NPRACH}$ (3 bits, *nprach-SubcarrierOffset*) ∈ {0, 12, 24, 36, 2, 18, 34} corresponds to the subcarrier offset within the 180kHz bandwidth.
4. $N_{sc}^{NPRACH}$ (2 bits, *nprach-NumSubcarriers*) ∈ {12, 24, 36, 48} corresponds to the number of subcarriers being used for random access.
5. $N_{start}^{NPRACH}$ (3 bits, *nprach-StartTime*) ∈ {8, 16, 32, 64, 128, 256, 512, 1024} in millisecond. This corresponds to the time of the start of the NPRACH transmission is $N_{start}^{NPRACH}$ subframes after the first subframe in radio frames fulfilling $mod\left(n_f, \frac{N_{period}^{NPRACH}}{10}\right) = 0$.

The NPRACH channel uses only 3.75KHz sub-carrier spacing and has a group of symbols (denoted as symbol group) each consisting of 5 symbols with a cyclic prefix (CP). Each NPRACH opportunity within an NPRACH resource is confined to 12 sub-carriers, which can span up to 128 repetitions. To reduce the relative CP overhead, $N$ sample OFDM symbol is repeated 5 times and a single CP of length $N_{cp}$ is added to form a symbol group. Two NPRACH Formats, i.e., Format 0 and Format 1 are defined for NB-IoT with $N_{cp}$ of 66.7 μs and 266.7 μs respectively. In this work the only NPRACH with preamble Format 0 is investigated, as shorter coverage areas are typically associated with small cells. To facilitate uplink timing estimation at eNB, the single tone preamble hops across different subcarriers. The hopping pattern consists of inner hopping fixed and outer pseudo-random hopping across repetitions (for any repetitions $N_{Rep}^{NPRACH}$ configured). Inner hopping is applied between the groups of 4 symbols groups with symbol groups 0,1 and 2,3 being one subcarrier apart; and symbol groups 1,2 being 6 subcarriers apart as shown in Figure 2-2. Outer hopping is across repetitions with a different start subcarrier index chosen from a pre-defined pseudo-random sequence (see [11]).

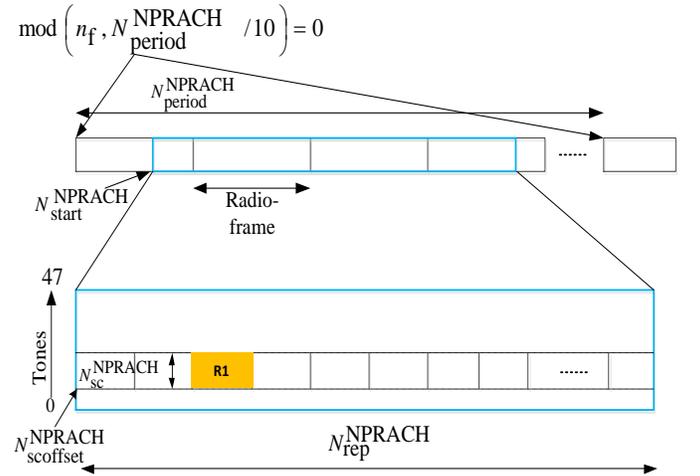

*Figure 2-1*: NPRACH time/frequency structure

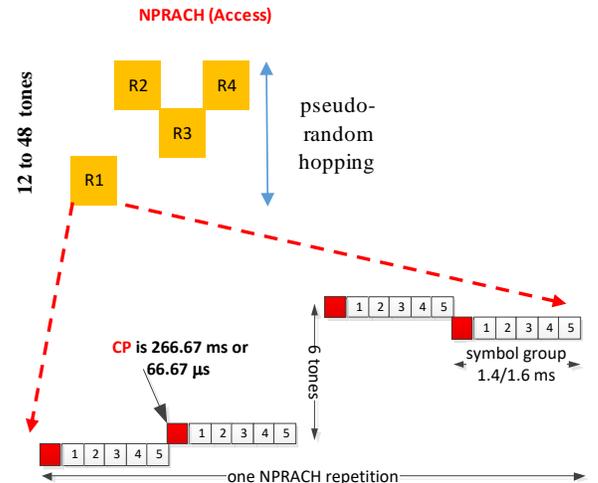

*Figure 2-2:* NPRACH inter and intra repetition hopping.

NB-IoT has the same OFDM symbol boundary as legacy LTE. That is, in a given subframe, symbol 0,7 have 128 data samples with 10 CP samples, and all the other symbols (i.e., 1 to 6 and 8 to 14) have 128 data samples with 9 CP samples with $f_s = 1.92$ MHz. There is no sampling rate below 1.92 MHz that can always align with the OFDM symbol boundary, i.e., have integer number of samples for both CP and data. This means that if lower input sampling rate is used, then time-domain interpolation is needed to adjust for the different sampling time offset. Sampling at 240kHz will require decimation by 4 with two finite impulse response filters. The complexity of processing samples at 1.92MHz or at 240kHz are very similar and the details are outside the scope of this paper. In this paper the sampling rate for PRACH is assumed to be $f_s = 1.92$ MHz. The derivation below is based on [5] and [9]. Let,

$$s_{m,i}[n] = e^{j2\pi n_{sc}^{RA}(m)n/N}, \quad \text{for } n = N_{m,i} - N_{cp}, \dots, N_{m,i} + N - 1 \quad (1)$$

be the transmit symbol at $n_{sc}^{RA}(m)$ subcarrier of $m^{th}$ symbol group with $N_{m,i} = mN_g + iN$, where, $s_{m,i}[n]$ – is the time domain waveform at $n^{th}$ sample of $i^{th}$ symbol in $m^{th}$ symbol group; $N_{cp}$ is size of CP (16 samples at $f_s = 240ksps$); $N$ is size of a symbol (64 samples at $f_s = 240ksps$); $N_g$ is size of a group ($N_{cp} + 5N = 336$ samples at $f_s = 240ksps = 1.4$ms). The received signal can be written as,

$$y_{m,i}[n] = h_{m,i} e^{j2\pi\xi(n-\tau)} s_{m,i}[n-\tau]$$
$$= h_{m,i} e^{j2\pi\xi(n-\tau)} e^{j2\pi n_{sc}^{RA}(m)(n-\tau)/N} \quad (2)$$

where $\xi$ is the frequency offset normalized with sampling frequency; $\tau$ is the round-trip delay (RTD) normalized with symbol duration; $h_{m,i}$ is the channel coefficient at the $i^{th}$ time sample of the $m^{th}$ symbol group and $j = \sqrt{-1}$. Dropping CP and taking FFT gives,

$$Y_{m,i}[k] = \sum_{n=N_{m,i}}^{N_{m,i}+N-1} y_{m,i}[n] e^{-j2\pi nk/N}. \quad (3)$$

When $k = n_{sc}^{RA}(m)$,

$$Y_{m,i} = \frac{h_{m,i} e^{j2\pi\xi(mN_g + iN - \tau)} e^{-j2\pi n_{sc}^{RA}(m)\tau/N}(1 - e^{j2\pi\xi N})}{1 - e^{j2\pi\xi}}. \quad (4)$$

There will be energy on other subcarriers as well. But the leakage is assumed to be small. Combining within a symbol group m, and assuming that the channel is invariant within a symbol group, we get (per antenna),

$$Y_m = \sum_{i=0}^{4} Y_{m,i}$$
$$= \frac{h_m e^{j2\pi\xi(mN_g - \tau)} e^{-j2\pi n_{sc}^{RA}(m)\tau/N}(1 - e^{j2\pi\xi 5N})}{1 - e^{j2\pi\xi}}. \quad (5)$$

Noise within a symbol group $m$ can be computed as,

$$N_m = |Y_{m,0} - Y_{m,1}|^2 + |Y_{m,3} - Y_{m,4}|^2 \quad (6)$$

The problem is to estimate the two unknowns, frequency offset $\xi$ and RTD $\tau$ using all available $Y_m$.

*A. Frequency offset estimation and correction*

First, $\tau$ is eliminated in equation (6) and the frequency offset $\xi$ is estimated by exploiting symbol combining across groups and repetitions (if configured). Towards this end, the following steps are performed. Define the product of symbol group $m$ with $m \in \{0,1,2,3\})$ and conjugate of symbol group $m + 1$ as,

$$Z_m = Y_m conj(Y_{mod(m+1,4)}) \propto e^{-j2\pi\xi N_g} e^{j2\pi H_m \tau/N} \quad (7)$$

Here, the hopping pattern $H_m$ is defined as,
$$H_m = n_{sc}^{RA}(m+1) - n_{sc}^{RA}(m)$$
$$\text{with } H_3 = n_{sc}^{RA}(0) - n_{sc}^{RA}(3)$$

Next combining across symbol groups within inner hopping (with $m \in \{0,1\}$) (see [9]), we get,

$$X_{m1} = Z_m + Z_{m+2} \propto e^{-j2\pi\xi N_g} cos(2\pi\tau/N).$$

Combining across repetitions (or over outer hopping) and number of receive antennas gives,

$$W_1 = \sum_{\substack{m \in \{0,4,\dots 4(R-1)\} \\ N_{rx}}} X_{m1} \propto e^{-j2\pi\xi N_g} \quad (8)$$

where $R$ is the NPRACH repetitions configured by upper layers. In the above equation, summation is also over number of receive antennas $N_{rx}$, and $R = N_{Rep}^{PRACH}$. Similarly, we can obtain,

$$X_{m2} = \begin{cases} Z_m - Z_{m+2} & H_m = +1 \\ Z_{m+2} - Z_m & H_m = -1 \end{cases}$$
$$\propto e^{-j2\pi\xi N_g} jsin(2\pi\tau/N)$$

$$W_2 = \sum_{\substack{m \in \{0,4,\dots 4(R-1)\} \\ N_{rx}}} X_{m2} \quad (9)$$
$$\propto e^{-j2\pi\xi N_g} jsin(2\pi\tau/N)$$

Note that computing $W_2$ may not be necessary if the frequency offset is not anticipated to be high. For example, with $\xi = 200$Hz, the angular rotation over a symbol group of size $N_g$ is $\approx 101$ degrees. For smaller values of $\xi$, computing only $W_1$ could suffice and could perhaps reduce computational complexity with marginal loss in performance. In order to make frequency offset estimation independent of $\tau$, we construct,

$$W = |W_1| \times W_1 - j|W_2| \times W_2$$
$$\propto e^{-j2\pi\xi N_g}(cos^2(2\pi\tau/N) + sin^2((2\pi\tau/N))$$
$$\propto e^{-j2\pi\xi N_g} \quad (10)$$

The frequency offset $\xi$, can then be estimated by computing the angle of $W$ as,

$$\xi = -\frac{1}{2\pi N_g} arctan(W). \quad (11)$$

Note that the maximum frequency offset which can be estimated with the above method corresponds to, $\xi_{max} = \pm \frac{1}{2N_g} = \pm \frac{1}{2*1.4ms} \approx 357Hz$, which is higher than the frequency offset setting of 200Hz for RAN4 specified NPRACH missed detection requirement (see table 8.5.3.2.1-1 of [14]). Note that the frequency offset need not be explicitly computed. The phaser corresponding to the frequency offset can be computed as,

$$\frac{W}{|W|} = e^{-j2\pi\xi N_g}. \quad (12)$$

Correcting the estimated frequency offset we get,

$$V_m = Z_m e^{j2\pi\xi N_g} \propto e^{j2\pi H_m \tau/N} \quad \text{and} \quad (13)$$
$$V_3 = Z_3 e^{-j2\pi 3\xi N_g} \propto e^{j2\pi H_3 \tau/N}$$

### B. Delay estimation

Defining,

$$U_{4r+l} = \begin{cases} V_{4r+l}, & if\ H_{4r+l} = H_l \\ V^*_{4r+l}, & if\ H_{4r+l} = -H_l \end{cases} \quad (14)$$

with $l \in \{-1, +1, 6\}$ the possible hopping sequences and summing across repetitions $r$, we get,

$$T_l = \sum_{r=0,1,\ldots R} U_{4r+l} \propto e^{j2\pi\alpha_l \tau/N} \quad (15)$$

Modify $T_l$ such that, $\alpha_0 = -1$, $\alpha_1 = 6$, $\alpha_2 = 1$ and $\alpha_3 = -6$. For a given NPRACH opportunity (parameterized by the subcarrier of the first symbol group $\rho = n_{sc}^{RA}(0)$) form vector $F_\rho = [-T_3\ 0\ 0\ 0\ 0\ T_0\ 0\ T_2\ 0\ 0\ 0\ 0\ T_1\ 0\ \ldots\ \ldots\ 0]$ with size $N_\tau$ being the FFT size used for $\tau$ estimation. The RTD can be estimated using ML estimator based on FFT. Let,

$$Q_v = \sum_{s=0}^{N_\tau - 1} F_\rho(s) e^{\frac{-j2\pi s}{N_\tau}} \quad \text{and}$$
$$G_v = \sum_{N_{rx}} |Q_v|^2 \quad (16)$$

The timing hypothesis vector $G_v$ corresponding to the maximum absolute value is the RTD estimate. Specifically,

$$[m_{max}, idx_{max}] = \underset{[0,N_\tau]}{\operatorname{argmax}}\{G_v\}. \quad (17)$$

RTD estimate can be computed using the FFT bin index ($idx_{max}$) corresponding to the maximum value $m_{max}$ as, $\tilde{\tau} = \frac{idx_{max}}{N_\tau*(3.75)kHz}$. Using $N_\tau = 256$ will give a resolution of $\frac{\frac{1}{(3.75)kHz}}{256} = 1.04\ us$. Given that the timing offset will be within the CP and the timing error is $\pm 7 \times 16 T_s$, the maximum is taken after zeroing out the FFT output corresponding to timing offsets outside the window determined by CP size and the timing error. A quadratic interpolation [16] around the maximum value provides better resolution in the timing estimate (but with additional computational complexity). Let the values of $G_v$ at $m_{max} - 1$, $m_{max}$ and $m_{max} + 1$ be $\alpha$, $\beta$ and $\gamma$ respectively as show in Figure 2-3. The maximum value after thresholding is given by,

$$G^* = y(p) = \beta - \frac{1}{4}(\alpha - \gamma) \times p,$$

where $p = \frac{\frac{1}{2}(\alpha - \gamma)}{\alpha - 2\beta + \gamma}$. RTD estimate can be refined after interpolation using $\tilde{\tau} = \frac{idx_{(max)} + p}{N_\tau*(3.75)kHz}$. For 128 repetitions[5], FFT at the output of two receive antennas and groups of 64 repetitions are non-coherently combined before determining the maximum.

---

[5] Based on the link budget analysis for NPRACH in small cells, high repetitions may not be necessary for the targeted MCL of 164dB.

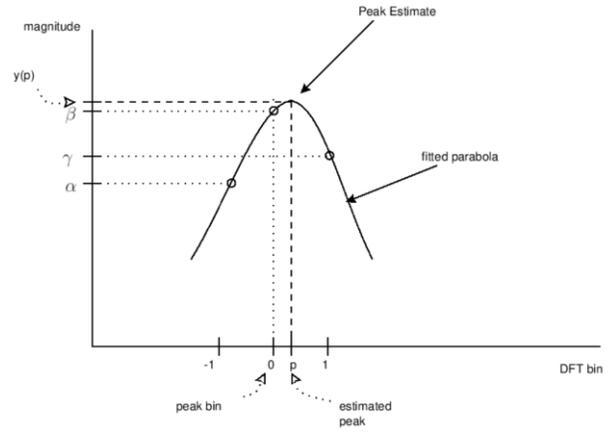

*Figure 2-3:* Polynomial interpolation around FFT bin with maximum value[16].

### C. Signal detection

NPRACH or DTX detection is determined based on comparing the metric $G^*$ normalized with the noise averaged across symbol groups, repetitions and antennas with a threshold $\epsilon_{NPRACH}^R$. Here $\epsilon_{NPRACH}^R$ is a constant determined experimentally by sending a zero signal (DTX) and computing the normalized value (normalized with noise power) of $G^*$ over several DTX frames. The cumulative distribution function (CDF) of the observed values of $G^*$ gives the probabilities of the values of $G^*$. The threshold $\epsilon_{NPRACH}^R$ is determined from the CDF by finding the value of $G^*$ occurring with a probability of 99.9% (corresponding to the false alarm rate of 0.1%). The detection on a subcarrier $\rho$ is done as follows.

$$\text{Let } \lambda = \frac{G^*}{\frac{1}{N_{rx} 4R} \sum_{r=0}^{R} \sum_{m=0}^{3} N_m}. \quad (18)$$

$$Detection\ on\ n_{sc}^{RA}(\rho)$$
$$= \begin{cases} NPRACH, & if\ \lambda \geq \epsilon_{NPRACH}^R \\ DTX, & otherwise \end{cases} \quad (19)$$

For the subcarrier $\rho$ on which NPRACH is detected, $SNR_\rho = \lambda$ can be reported to upper layers for scheduling/uplink power control algorithms.

### 3. NPUSCH FORMAT 1 (DATA) RECEIVER

NB-IoT UL shared channel data is mapped to NPUSCH Format 1 and supports 1,3,6, or 12 tones at 15 kHz subcarrier spacing or a single tone at 3.75 kHz subcarrier spacing. Since the bandwidth is limited to 180 kHz, resource units (RU) are defined, which are essentially resource blocks allocated in time. Details on resource unit duration in millisecond (ms), slot lengths in ms, modulation used, number of symbols and references signals are shown for NPUSCH Format 1 in Figure 3-1. NPUSCH Format 1 has one pilot symbol per slot, uses Turbo coding and uses either $\frac{\pi}{4}$ QPSK or $\frac{\pi}{2}$ BPSK modulation.

NPUSCH Format 1 is used to carry uplink data, which is processed in the form of transport blocks (TB).

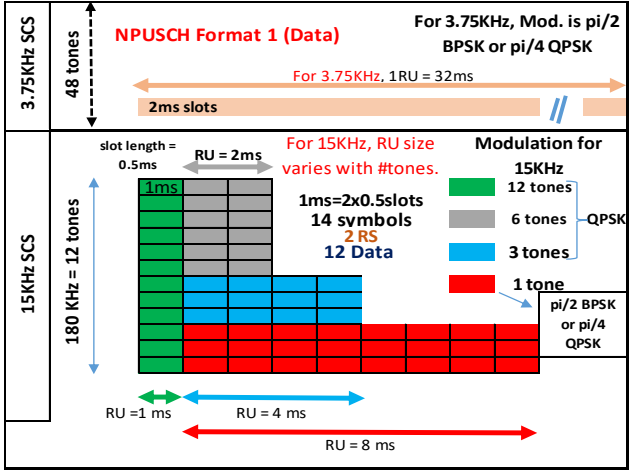

Figure 3-1:NPUSCH Format 1 (Data) Resource Unit Structure.

A transport block is encoded as one coding block and transmitted $N_{rep}^{NPUSCH}$ times contiguously, unless uplink transmission gaps are required. Each of the $N_{rep}^{NPUSCH}$ repetitions is mapped to $N_{RU}$ resource units (RU), where $N_{RU}$ may be one of the values in {1, 2, 3, 4, 5, 6, 8, 10} depending on the TBS size and MCS index (see section 16.5.1.2 of [13] for TBS table). A resource unit is defined as $N_{symb}^{UL} N_{slots}^{UL}$ consecutive SC-FDMA symbols in the time domain and $N_{SC}^{RU}$ consecutive subcarriers in the frequency domain, where $N_{SC}^{RU}$ and $N_{symb}^{UL} = 7$ are the number of assigned tones and number of symbols in a slot respectively (see [11]). The $N_{rep}^{NPUSCH}$ repetitions are grouped into $\frac{N_{rep}^{NPUSCH}}{M_{identical}^{NPUSCH}}$ repetition cycles. Within each cycle, every $N_{slots}^{UL}$ are repeated $M_{identical}^{NPUSCH} - 1$ identical times before the next $N_{slots}^{UL}$ are transmitted, where $M_{identical}^{NPUSCH} = \min\left(\frac{M_{rep}^{NPUSCH}}{2}, 4\right)$ if $N_{SC}^{RU} > 1$ and $M_{identical}^{NPUSCH} = 1$ if $N_{SC}^{RU} = 1$.

Consider the $N_l$ samples from $s^{th}$ slot and the $l^{th}$ symbol i.e. in the range $n = N_{s,l} - N_{cp}^l, \ldots, N_{s,l} + N_l - 1$. Here $N_{s,l}$ are the samples up to this symbol given by, $N_{s,l} = sN_{slot} + \sum_l(N_{cp}^l + N_l)$. Let, $x_{s,l}[n] = \alpha_l e^{\phi_{k,l}} e^{\frac{j2\pi kn}{N_l}}$ is the transmit signal on the $k^{th}$ subcarrier with modulated alphabet $\alpha_l$. Here, $N_{slot}$ is the size of a 0.5ms slot[6]. $N_l$ is the size of a symbol, with $l \in \{0, \ldots, 6\}$. $N_{cp}^l$ is the cyclic prefix[7] size of the $l^{th}$ symbol. $x_{s,l}[n]$ is the time domain waveform at $n^{th}$ sample of $l^{th}$ symbol in $s^{th}$ slot. Also, $\phi_{k,l}$ is the phase of the $k^{th}$ subcarrier.

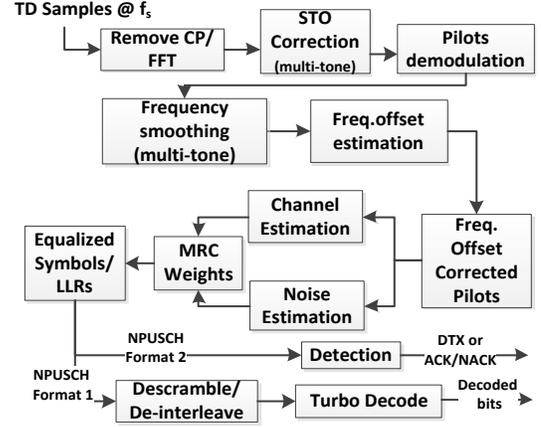

Figure 3-2:Receiver blocks for NPUSCH F1 and F2.

To minimize the phase discontinuity between symbol boundaries, a modulation-based phase rotation is applied (see section of 10.1.5 of [10]). For symbol l of a continuous transmission over subcarrier $k$, the modulation-based phase rotation to be applied is $e^{j\phi_{k,l}}$. The modulation-based phase rotation essentially consists of two parts, first part is the $\frac{\pi}{2}$ or $\frac{\pi}{4}$ phase rotation for BPSK or QPSK modulation respectively which is applied to odd-numbered symbols, second the accumulative phase rotation part that is used to ensure the desired phase difference between symbol boundaries. This phase rotation is accumulative throughout a contiguous transmission. The received signal for slot $s$ and symbol $l$ can be written as,

$$y_{s,l}[n] = h_{s,l}e^{j2\pi\xi(n-\delta t)}x_{s,l}[n-\delta t] + w_{s,l}[n]$$
$$= h_{s,l}e^{j2\pi\xi(n-\delta t)}\alpha_l e^{j\phi_{k,l}}e^{\frac{j2\pi k}{N_l}(n-\delta t)} + w_{s,l}[n]. \quad (20)$$

where $\xi$ is the frequency offset between the UE and eNB transmit/receive frequencies normalized with sampling frequency; $\delta t$ is the sampling time offset normalized with symbol duration; $h_{s,l}$ is the channel coefficient at the $l^{th}$ symbol of $s^{th}$ slot; $w_{s,l}$ is the white Gaussian noise at the $s^{th}$ slot and $l^{th}$ symbol and $j = \sqrt{-1}$. Dropping the CP and taking FFT, we get equation frequency domain samples as,

$$Y_{s,l}[k] = \sum_{n=N_{s,l}}^{N_{s,l}+N_l-1} y_{s,l}[n]e^{-\frac{j2\pi nk}{N_l}}. \quad (21)$$

Simplifying further we get (adding $\delta t$ notation to indicate dependence on timing offset[8]), the frequency domain sample for slot $s$, symbol $l$, on the $k^{th}$ tone as,

---

[6] This is equal to 960 samples at sampling rate of $f_s = 1.92\ Msps$.

[7] At $f_s = 1.92\ Msps$, this corresponds to $N_{cp}^l = 10\ for\ l = 0$ and $N_{cp}^l = 9\ for\ l > 0$.

[8] The frequency domain samples of the received signal are also a function of the frequency offset $\xi$, but the notation is dropped to avoid clutter.

$$Y_{s,l,k}^{\delta t} = \frac{h_{s,l,k} \alpha_l e^{j\phi_{k,l}} e^{j2\pi\xi(N_{s,l}-\delta t)} e^{\frac{-j2\pi k \delta t}{N_l}} (1 - e^{j2\pi\xi N_l})}{(1 - e^{j2\pi\xi})} + W_{s,l,k}. \quad (22)$$

### A. Timing offset estimation for multi-tone assignments:

In general, the received signal is not time aligned and typically has a non-zero sampling time offset. Timing offset estimation is needed in the receiver because different tones have different phases which can generally be captured by the channel estimate as well. However, to enhance performance, it is better to separate this offset from channel estimate. In the case of single tone, this is not needed because there is no second tone, so the problem of different phases between tones does not even arise. Computing the phase difference across adjacent subcarriers $k$ and $k+1$, we get,

$$P_k = Y_{s,l,k}^{\delta t} * \text{conj}(Y_{s,l,k+1}^{\delta t}) \propto e^{\frac{j2\pi\delta t}{N_l}}. \quad (23)$$

Summing the phase differences across subcarriers for a give slot $s$, we get $Q_k^s = \sum_{k=1}^{N_{sc}-1} P_k$. The sampling time offset for a given slot can be estimated as,

$$\widetilde{\delta t} = \frac{1}{2\pi N_l} \arctan(Q_k^s). \quad (24)$$

The average sampling time offset across $n$ slots can be used for final sampling time offset correction. Note that after the initial NPRACH access, UE would apply the eNB suggested timing advance (within a $\pm 3.43 \mu s$ accuracy [14]) to achieve uplink synchronization and whatever subsequent sampling time offset which is seen by eNB is either due to the residual timing offset or due to the drift in the UEs local oscillator over time. If the UE timing off by a large value, an NPDCCH order is typically sent to allow the UE to reconnect. Note that sampling time offset need not be computed explicitly, and the sampling time offset correction on the received tones can be done using the phasor, $\frac{Q_k^s}{|Q_k^s|} = e^{\frac{j2\pi \widetilde{\delta t}}{N_l}}$. The timing correction can be applied as $Y_{s,l,k} = Y_{s,l,k}^{\delta t} * \frac{Q_k^s}{|Q_k^s|}$.

### B. Pilot Demodulation:

The first step is pilot demodulation i.e., removal of known values of the pilots from the received pilot signal by a multiplication of the signal with the conjugated version of the known pilot sequence. For $k^{th}$ tone of the $l^{th}$ symbol in slot $s$ (all within a block $b$), pilot demodulation can be written as,

$$\hat{h}_{s,l,k} = r_{s,l,k}^* Y_{s,l,k} \quad (25)$$

Detailed information on generation of the known pilot sequence $r_{s,l,k}$ can be found in [S10]. Let the transmitted pilot signal be $r_{s,l,k}$, the frequency offset be $\xi$ (normalized with sampling period) and noise be $n_{s,l,k}$, then the received pilot signal (assuming that the received signal is time offset corrected) can be written as $Y_{s,l,k} = h_{s,l,k} r_{s,l,k} e^{j2\pi\xi N_{s,l}} + n_{s,l,k}$. Multiplying with the complex conjugate $r_{s,l,k}^*$ results in $h_{s,l,k} e^{j2\pi\xi N_{s,l}} +$ $m_{s,l,k}$, since $|r_{s,l,k}|^2 = 1$. An FFT based frequency offset estimator is generally used on such phase ramped signals. For NPUSCH Format 1, $l = 3$ is the pilot symbol. For multi-tone transmissions, frequency smoothing is first applied to the demodulated pilot. A linear average of all tones in each of the received SC-FDMA symbol it taken to increase SNR under relatively flat channel conditions. Specifically,

$$\hat{h}_{s,3} = \frac{1}{N_{SC}^{RU}} \sum_{k=0}^{N_{SC}^{RU}-1} \hat{h}_{s,3,k} \quad (26)$$

Here $N_{SC}^{RU}$ is the number of subcarriers of the resource unit assigned to the user under consideration. The averaged signal is used for channel estimation, frequency offset estimation and for noise power estimation. The pilot symbols (after conjugate matching and averaging because of flat channel in the narrow band) form a noisy phase ramp signal with a sampling period of 0.5 ms.

### C. Frequency offset estimation:

For frequency offset and channel estimation a block-based processing scheme is adopted to avoid large memory requirements during higher repetitions and to obtain processing gain during low operating SNR. Block sizes of $B = 8\ ms$ is used and block size of $B = 32\ ms$ for multi and single tone assignments respectively, when a single transmission time are greater than $B$, otherwise the transmission time is used as the block size. The block size is chosen such that the RAN4 performance targets are met. The block size is chosen to be smaller than the channel coherence time $\approx \frac{0.423}{f_d} = 423$ ms for $f_d = 1 Hz$ Doppler. With pilots occurring every 0.5ms and using a $N = 256$ point FFT, we get a resolution of $2000\ Hz/256 \approx 7.8\ Hz$ in frequency offset $\xi$. The output of the FFT is squared and summed over the receive antennas and the index corresponding to the maximum value provides an estimate of the frequency offset as shown in equation (26).

$$H_v = \sum_{m=0}^{N-1} \hat{h}_{s,3} e^{\frac{-j2\pi m}{N}}$$

$$I_v = \sum_{N_{rx}} |H_v|^2 \quad (27)$$

$$m_{max} = \underset{\left[-\frac{N}{4}, \frac{N}{4}\right]}{\arg\max} \{I_v\}$$

The maximum is taken over a range of possible frequency offset values of $[-250\ 250] Hz$. Further accuracy in the estimate can be obtained by interpolating around the maximum value (like what was defined in NPRACH receiver section 2.B) as follows. Let the values of $G_v$ at $m_{max} - 1$, $m_{max}$ and $m_{max} + 1$ be $\alpha$, $\beta$ and $\gamma$ respectively. The offset $p$ of the true peak w.r.t $m_{max}$ is given by, $p = \frac{\frac{1}{2}(\alpha - \gamma)}{\alpha - 2\beta + \gamma}$. The frequency offset estimate $\hat{\xi}$ is given by, $\hat{\xi} = \frac{m_{max} + p}{NN_{slot}}$.

*D. Channel estimation:*

The channel estimation process is repeated every block for each active UE. The channel estimates are averaged over the block. This design follows the assumption of a slow fading channel ($\approx$ 1Hz Doppler frequency). The time averaging operation is done together with the frequency offset correction operation for antenna $r$ as,

$$\hat{h}_{b,r} = \frac{1}{2B} \sum_{s=0}^{2B-1} h_{s,3} e^{-j2\pi\hat{\xi}N_{s,3}}. \quad (28)$$

For multiple receive antennas, $\hat{h}_{b,r}$ is a vector of size $N_{rx}$ with elements corresponding to channel estimates at each of the receive antennas. Note that the subscript $b$ denotes a given block. The frequency offset corrected data symbols for antenna $r$ are given by,

$$d_{s,l,k,r} = Y_{s,l,k} e^{-j2\pi\hat{\xi}N_{s,l}} \quad (29)$$

The noise variance is computed as shown below.

$$\sigma_b^2 = \frac{1}{2BN_{rx}N_{sc}^{RU}} \sum_{N_{rx}} \sum_{N_{sc}^{RU}} \sum_{s=0}^{2B-1} |\hat{h}_{s,3,k,r} e^{-j2\pi\hat{\xi}N_{s,3}} - \hat{h}_{b,r}|^2 \quad (30)$$

Note that there are two slots per subframe and hence the summation is over $2B$ for a block of size $B$ ms. When the block size $B$ is small, the number of samples in the noise estimate is too small and the estimates could be biased. In such a scenario, the SNR could be estimated as,

$$\sigma_b^2 = \frac{1}{2BN_{rx}N_{sc}^{RU}} \sum_{N_{rx}} \sum_{N_{sc}^{RU}} \sum_{s=0}^{2B-1} |\hat{h}_{s,3,r} e^{-j2\pi\xi N_{s,3}} - \hat{h}_b|^2 \quad (31)$$

Here $\hat{h}_b = \frac{1}{2B}\sum_{s=0}^{2B-1} h_{s,3} e^{-j2\pi\xi N_{s,3}}$ (time averaged) and $\hat{h}_{s,3,r} = \frac{1}{N_{sc}^{RU}}\sum_{k=0}^{N_{sc}^{RU}-1} \hat{h}_{s,3,k}$ (frequency averaged). Also, $\hat{h}_{s,3,k}$ is the demodulated, STO corrected (multi-tone) reference symbol in a given slot $s$ and for a given tone $k$. Under MMSE criteria, equalizer taps for block $b$ and antenna $r$ are calculated in the following manner,

$$w_{b,r} = \sigma_b^{-2} \hat{h}_{b,r}^H, \quad (32)$$

with $(\cdot)^H$ denoting the Hermitian operation. The equalized symbols in a block $b$ are given by,

$$e_{s,l,k} = \sum_{N_{rx}} w_{b,r} d_{s,l,k,r}. \quad (33)$$

where the subscript $r$ denotes the receive antenna. The inverse Fourier transform of order $N_{sc}^{RU} \in \{3,6,12\}$ is computed as,

$$e_{s,l,k}^t(n) = \frac{1}{\sqrt{N_{sc}^{RU}}} \sum_{k=0}^{N_{sc}^{RU}-1} e_{s,l,k} e^{-\frac{j2\pi nk}{N_{sc}^{RU}}} \quad (34)$$

The log-likelihood ratios (LLRs) are obtained as,

$$llr_{s,l,k} = \begin{cases} -Real(e_{s,l,k}^t) - Imag(e_{s,l,k}^t) & \text{for } BPSK \\ \left(-Real(e_{s,l,k}^t), -Imag(e_{s,l,k}^t)\right) & \text{for } QPSK \end{cases}$$

The LLRs corresponding to identical repetitions (i.e., LLRs belonging to the same resource units) are combined before descrambling and de-interleaving. Rate de-matching is performed for the de-interleaved LLRs for each repetition and the repetitions of the same transmission are combined with HARQ LLRs of earlier transmissions if available. For every non-identical repetition, a turbo decoding operation is performed followed by a CRC check. Bits outputs from the decoder after successful decoding, including CRC check indication for each identical repetition group are transferred to higher layer. The SNR over a block $b$, is given by,

$$SNR_b = w_b \hat{h}_b. \quad (35)$$

The SNR reported to higher layers is generally the average SNR over the last few blocks to avoid long time averaging for higher repetitions.

## 4. NPUSCH FORMAT 2 (CONTROL) RECEIVER

The NPUSCH Format 2 channel carries acknowledgement of downlink data and supports only single tone with either 3.75 kHz or 15 kHz subcarrier spacing. The NPUSCH Format 2 channel has a resource unit (RU) duration of 2 ms and has three reference symbols per slot to facilitate robust estimations while using a single 3.75 kHz/15 kHz tone. In this work, only the 15 kHz sub-carrier spacing is considered for NPUSCH Format 2. The single tone NPUSCH Format 2 is similar to the single tone NPUSCH Format 1, except for the pilot structure, channel coding and modulation scheme.

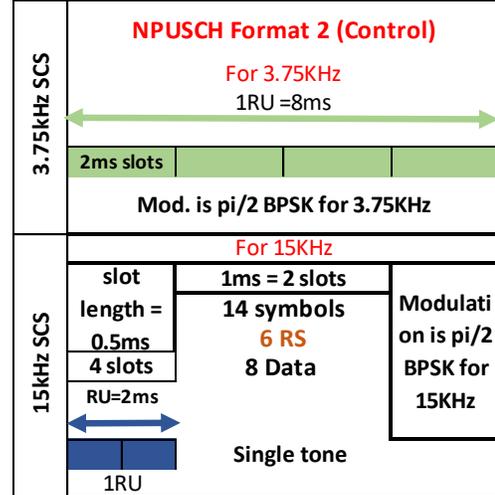

**Figure 4-1: NPUSCH Format 2 (Control) Resource Unit Structure.**

NPUSCH Format 2 has three pilot symbols per slot, uses repetition coding on one ACK/NACK bit and uses only a $\frac{\pi}{2}$ BPSK modulation scheme. The frequency offset estimation and channel estimation modules proposed in previous section for NPUSCH Format 1 can be reused for NPUSCH Format 2 also. One minor change, however, is that in case of NPUSCH Format 2, there are three pilots available in a slot of 0.5ms duration compared to a single pilot in case of NPUSCH Format 1. The three pilots are time averaged, assuming minor phase changes between adjacent symbols. Pilot demodulation can then proceed as in NPUSCH Format 1. Reusing NPUSCH Format 1 modules has the advantage of saving memory in an implementation, reduction in development and module testing efforts. For NPUSCH Format 2, there is no decoding of the LLRs involved and only the ACK/NACK detection has to be designed, as discussed below. Let $\alpha_l = t_l u$ be the modulated alphabet transmitted on sub-carrier $k = k_0$, where $t_l$ is the scrambling and $u$ is the unknown ACK/NACK bit. For every symbol, compute,

$$Z_{s,l}[k_0] = Y_{s,l}[k_0] t_l^* e^{-j\phi_{k_0,l}}, \tag{36}$$

where $Y_{s,l}[k_0]$ is defined in equation (21) and, $e^{j\phi_{k_0,l}}$ is the modulation-based phase rotation to be applied for symbol $l$ of a continuous transmission over subcarrier $k_0$ (see section of 10.1.5 of [10]). Dropping the subcarrier notation $k_0$ to avoid clutter, for slot $s$, we define variables $Z_s^p$, and $Z_s^d$ for pilot and data symbols respectively for each antenna as,

$$Z_s^p = \sum_{l=2,3,4} Z_{s,l} \propto h_g e^{j2\pi\bar{\xi}N_{s,l}}$$
$$Z_s^d = \sum_{l=0,1,5,6} Z_{s,l} \propto h_g u e^{j2\pi\bar{\xi}N_{s,l}}$$

$$J_r(u) = \sum_{s=gP+1}^{(g+1)P} (Z_s^p + u Z_s^d) w_b e^{-j2\pi\bar{\xi}N_{s,l}} \tag{37}$$

$$J(u) = \sum_{r=1}^{N_{rx}} |J_r(u)|^2.$$

Here, $\bar{\xi}$ is the estimated frequency offset (as shown in section 3.C), $w_b$ is the equalizer tap weight computed as in equation (32) but with noise variance computed differently as shown in equation (39) of section 4.A. Also, $g$ represents the resource unit (RU) number, $P$ is the number of repetitions configured and $N_{rx}$ is the number of antennas. The sign of the equalized symbol and the SNR of the transmission $SNR_t$, which is the sum of the SNRs of all blocks in the transmissions, i.e., $SNR_t = \sum_B SNR_b$, are used as below to determine whether ACK, NACK or DTX was transmitted,

$$\frac{A}{N} = \begin{cases} DTX, & \text{if } SNR_t \leq \epsilon \\ sign(\underset{u}{argmax}[Re(J(u))]), & \text{otherwise.} \end{cases} \tag{38}$$

where, $\frac{A}{N} \in \{1, 0, -1\}$ represents the ACK/NACK or DTX information respectively. Here, $SNR_b$ is computed according to equation (35), $\epsilon$ is a constant determined by sending a zero signal (DTX) and computing the value of $SNR_t$ over several DTX frames. CDF of the observed values of $SNR_t$ is plotted to determine the threshold $\epsilon$ which gives 1% DTX to ACK.

*A. Noise estimation for NPUSCH F2:*

Since $N_{SC}^{RU} = 1$ (i.e., single tone) for NPUSCH Format 2 and the block size $B = 4ms$, the noise estimation used in NPUSCH Format 1 may lead to biased estimates during low SNRs due to insufficient averaging. Exploiting the fact that for NPUSCH Format 2 all data symbols are identical i.e., either ACK or NACK bit, the differences between adjacent symbols will unmask the noise. With this approach, 4 samples can be obtained per millisecond per antenna and averaging over $B = 4ms$ gives better estimation accuracy. Specifically, noise can be estimated as,

$$\sigma^2 = \frac{1}{2BN_{rx}} \sum_{N_{rx}} \sum_{\substack{s=0 \\ l=0,1,5,6}}^{2B-1} |d_{s,l,k_0} - d_{s,l+1,k_0}|^2 \tag{39}$$

where $d_{s,l,k_0}$ is the unscrambled data symbol $l$ corresponding to $s^{th}$ slot transmitted on sub-carrier $k_0$. Note that there are two slots per subframe and hence the summation is over $2B$ for a block of size $B$ ms.

5. IMPLEMENTATION ASPECTS

In a typical eNB implementation, concurrent processing of legacy LTE, eMTC, NB-IoT uplink channels, is usually required with multiple users being served per transmit time interval (TTI). Therefore, it will be useful to discuss the various implementation aspects associated with NB-IoT uplink receiver. In this section some of the important implementation aspects are discussed.

*A. Common Phase Correction:*

In a practical setup, phase correction per symbol is required to account for the difference between frequency location of the center of the Narrowband IoT physical resource block (PRB) and the frequency location of the center of the LTE signal, since a single common wideband receiver FFT (of say size 2048 for 20MHz bandwidth at 30.72Msps) is typically used in the digital front end (DFE) for both legacy LTE and NB-IoT. Due to the common FFT, there will be a phase offset introduced for NB-IoT tones due to the difference between center frequencies of legacy and NB-IoT. Consider the multi-tone NB-IoT signal $s(t)$ (generated according as in section 10.1.5 in [11]),

$$s(t) = \sum_{k=-6}^{5} a_k e^{j2\pi\left(k+\frac{1}{2}\right)\Delta f(t-N_{cp,l}T_s)} \tag{40}$$

See Appendix A for symbol definitions. The origin in the above equation is at the beginning of the $l^{th}$ OFDM symbol (including cyclic prefix). Moving it to the beginning of the $l^{th}$ OFDM symbol without CP, we get

$$s(t) = \sum_{k=-6}^{5} a_k e^{j2\pi\left(k+\frac{1}{2}\right)\Delta f t}. \tag{41}$$

If the separation between the centers of LTE and NB-IoT carriers is $f_o = N_o \Delta f$, where $\Delta f$ is the sub-carrier spacing and $N_0$ is some constant, the received signal is given by,

$$s(t) = \sum_{k=-6}^{5} a_k e^{j2\pi\left(k+\frac{1}{2}\right)\Delta f t} e^{j2\pi f_o(t+M_o T_s)}. \tag{42}$$

Here, $M_o T_s$ is the time offset between beginning of frequency shift application and the $l^{th}$ OFDM symbol (excluding CP) under consideration. Before taking FFT, the signal must be frequency shifted by $\pm\frac{1}{2}\Delta f$ (half tone shift) so that the FFT output has some scaled version of the transmit tones (instead of frequency interpolated samples),

$$s(t) = \sum_{k=-6}^{5} a_k e^{j2\pi\left(k+\frac{1}{2}\right)\Delta f t} e^{j2\pi\left(f_o \pm \frac{1}{2}\Delta f\right)(t+M_o T_s)} \quad (43)$$

Sampling gives,

$$s_n = \sum_{k=-6}^{5} a_k e^{j2\pi\left(k+\frac{1}{2}\right)\Delta f n T_s} e^{j2\pi\left(f_o \pm \frac{1}{2}\Delta f\right)(nT_s+M_o T_s)} \quad (44)$$

Taking FFT results in,

$$S_m = \frac{1}{N}\sum_{n=0}^{N-1} s_n e^{-\frac{j2\pi mn}{N}} \quad (45)$$

$$= \frac{1}{N}\sum_{n=0}^{N-1}\sum_{k=-6}^{5} a_k e^{j2\pi\left(k+\frac{1}{2}\right)\Delta f n T_s} e^{j2\pi\left(f_o \pm \frac{1}{2}\Delta f\right)(nT_s+M_o T_s)} e^{-\frac{j2\pi mn}{N}}$$

$$S_{k+N_o+\frac{1}{2}\pm\frac{1}{2}} = e^{j2\pi\left(N_0 \pm \frac{1}{2}\right)M_o T_s \Delta f} a_{k+\frac{1}{2}\pm\frac{1}{2}}$$

$$= e^{\frac{j2\pi\left(N_0 \pm \frac{1}{2}\right)M_0}{N}} a_{k+\frac{1}{2}\pm\frac{1}{2}}.$$

The phase correction to be applied at the NB-IoT receiver is then $e^{-\left(\frac{j2\pi\left(N_0 \pm \frac{1}{2}\right)M_0}{N}\right)}$.

*B. Uplink Transmission Gap Handling:*

The frequency offset estimation and channel estimation and identical LLR combing modules will be impacted in the presence of UL transmission gaps. From section 10.1.3.6 of [10], any ongoing NPUSCH transmissions will be postponed by the UE when an NPRACH resource overlaps in time and frequency and will be resumed after the NPRACH opportunity is complete. Also, to maintain time/frequency synchronization between UE and eNodeB during long UL repetition transmissions, UL gaps can be created. During UL gaps, the UE may switch to the DL and performs time/frequency synchronization. If an NPUSCH UL transmission duration is ≥ 256 ms[9], then a gap of 40 ms is inserted by the UE, followed by the remaining transmissions. Further, UL gaps can be created due to base-station configuring invalid subframes during which NB-IoT transmission cannot take place [13][17]. Such UL transmission gaps will require special handling by the receiver due to channel changing over the UL gap. There are several ways of handling this scenario. However, the simplest method could be to stop the block processing (for frequency offset and channel estimations) with a block of size ≤ to 8/32 ms. The processing can be resumed after NPRACH transmission with a block of size ≤ 8/32 ms. From link level simulations, it was observed that the impact on the performance with the above proposed method was marginal.

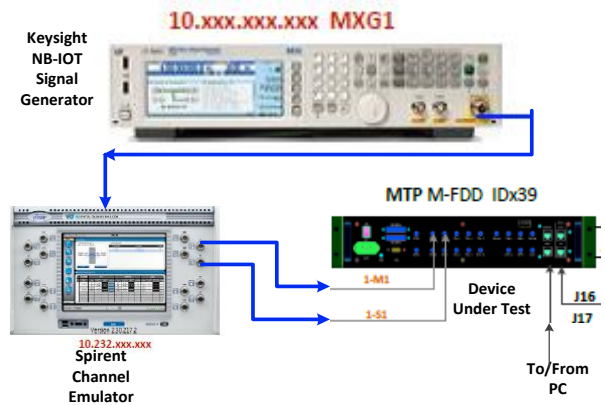

*Figure 6-1:* Test Setup

6. PERFORMANCE RESULTS AND DISCUSSIONS

In this section results are provided with link-level simulations of all uplink channels and discussions are provided on the NB-IoT receiver performance on a Qualcomm FSM Chipset based LTE basestation. The test setup with which the performance of NB-IoT uplink receiver was investigated is shown in Figure 6.1. The practical setup was shown in Figure 6.1 consisted of a keysight MXG signal generator capable of generating NB-IoT uplink signals[19], a spirent channel emulator [20] box capable of generating 3GPP specified channel profiles (e.g., EPA channel with 5Hz Doppler and low correlation between antennas), and the device under test (shown as MTP in the Figure 6.1). The test procedure was performed as per [15] as follows. For the given NB-IoT Channel under test, do the following.

**Step 1)** Adjust the AWGN generator, according to the channel bandwidth (see section 8.5 of [15]. **Step 2)** The characteristics of the wanted signal shall be configured according to the corresponding UL reference measurement channel defined in Annex A of [15]. **Step 3)** The multipath fading emulators shall be configured according to the corresponding channel model defined in Annex B of [15]. **Step 4)** Adjust the frequency offset (if defined) of the test signal.

| $N_{Tx}$ | $N_{RX}$ | $N_{rep}$ | Channel | FO $\xi$ | SNR[dB] | | |
|---|---|---|---|---|---|---|---|
| | | | | | SNR at which min detect. is req. [14] Preamble Format 0 | NPRACH Performance for Format 0 with Simulation | NPRACH Performance for Format 0 with Test Setup |
| 1 | 2 | 8 | AWGN | 0 | -2.1 | -4.5 | -4.1 |
| | | | EPA1 Low | 200 Hz | 6.1 | 3.65 | 4.12 |
| | | 32 | AWGN | 0 | -6.8 | -9.23 | -8.5 |
| | | | EPA1 Low | 200 Hz | 0.5 | -1.75 | -1.0 |

*Table 1: NPRACH RAN4 performance summary.*

[9] For NPRACH, if the transmission duration is ≥ 64 × preamble duration, then a gap of 40ms is inserted at the end of the NB-PRACH opportunity.

**Step 5)** Adjust the equipment so that the SNR specified in Tables 8.5.x of [15] is achieved at the BS input. **Step 6)** The test signal generator sends a test pattern (outlined in section 8.5 of [15]) with repetitions and the receiver tries to detect the pattern. This pattern is then repeated. For each of the reference channels (defined in Table 8.5.x of [15]) applicable to base station, the throughput is measured (as per Annex E of [15]) for NPUSCH Format 1 testing. For NPUSCH Format 2 testing, the statistics of ACK/NACK/DTX is noted as per section 8.5.2.4.2 of [15]. For NPRACH, the statistic are collected as per section 8.5.3.4.2 of [15]. Only the RAN4 recommended tests [14] were studied in both simulations and the test setups.

The RAN4 test results for NPRACH is summarized in Table 1 (see page 10). The SNR in dB column has three columns under it. The first colum has the 3GPP specified minimum SNR at which the probability of preamble detection is greater than or equal to 99% with detection requirements as specified in section 8.5.3.2 of [14] and with false alarm probability being less than or equal to 0.1% (see section 8.5.3.1 of [14]). The second column under SNR captures the NPRACH performance with link simulations and the third column lists the NPRACH performance on the test setup. As can be seen, there is a margin of atleast >2dB ($\approx$ 1.5 db) with link-level simulations and the test setup over the RAN4 recommended SNR. This demonstrates that the proposed NPRACH receiver comfortably meets the RAN4 minimum requirements. Note that, becaue of the multiplication between sum and difference differentials in equation (10), the noise in the differentials will have a multiplicative effect and the performance could be poorer with AWGN channel or when the frequency offset is low. Using $W = W_1$ (in equation (10)) during such a scenario could improve NPRACH performance. However, with NB-IoT devices expected to be designed with low-cost hardware, there could be a large frequency offset between the transmitter and the receiver, and the step in equation (10) might become necessary for better estimation of $\xi$.

The NPRACH performance obtained with the proposed design is comparable with the performance reported in [22]. The differences in link-level simulations and FSM mplementation (i.e., with test setup) is mainly due to the loss from the analog receive filter which was not modelled in the link level simulations, and the fixed point implemenation losses (<0.5dB). Note that other impairments like antenna correlation, I/Q imbalance, phase noise, etc which were not modelled in the link level simulations, but could be part of the test setup. Further, cabelling and signal splitter losses (within the Spirent channel emulator shown in Figure 6.1) add to the losses in test setup.

The performance of NPUSCH Format 1 (data) for single and 6-tones are shown in Figure 6.2 and Figure 6.3 respectively. Table 2 lists NPUSCH Format 1 test results (see page 11) for all RAN4 specified (see section 8.5.1 of [14]) test cases. As per RAN4 [14], for single tone allocation, minimum performance is specified only for repetitions of 1, 16 and 64 (indicated by diamond markers in Figure 6.2), whereas for multi-tone allocation, minimum performance is specified only for repetitions of 2, 16 and 64 (denoted by diamond markers in Figure 6-3). The solid lines (black, red and blue) in Figures 6.2 and 6.3 indicate the performance obtained with link level simulations. The performance requirement of NPUSCH Format 1 is determined by a minimum required throughput for a given SNR with HARQ retransmissions. The required throughput is expressed as a fraction (70%) of the maximum throughput. The link-level simulation plots for the normalized throughputs (normalized with the maximum throughput) are shown for different NPUSCH Format 1 repetitions in Figure 6.2. It can be seen that the demodulation performance improves (i.e., with higher margins) with higher repetitions, but incurs higher delays in detection/demodulation. However, for Small Cells, with coverage requirements being smaller than macro basestations [1], higher repetitions may rarely be configured. The summary of simulation and test setup results for all RAN4 specified test cases with NPUSCH Format 1 using 15kHz subcarrier spacing (Table 8.5.1.1.1-3 of [14]) is captured in Table 2 (see next page). The results are in line with the performance reported in [23].

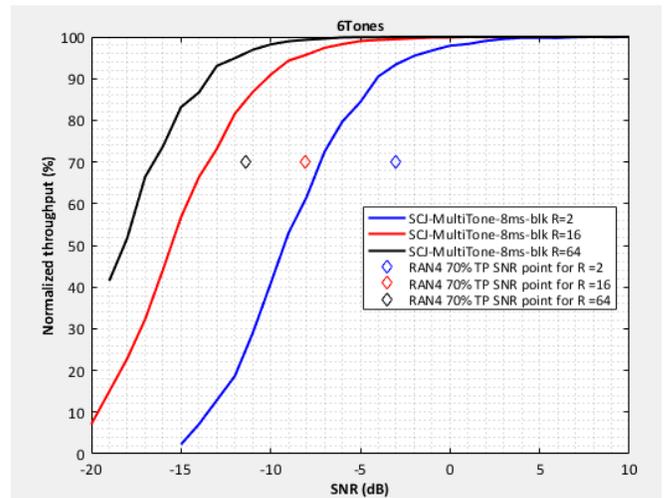

*Figure 6-2:* Link Performance of NPUSCH F1 for single tone

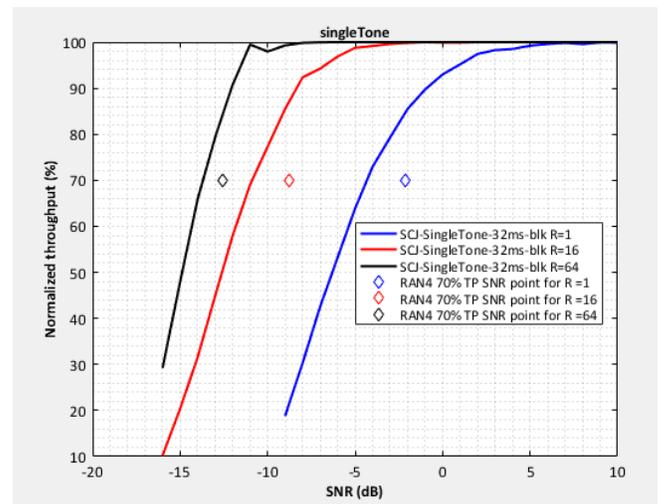

*Figure 6-3:* Link Performance of NPUSCH F1 for 6 tones

The minimum performance of NPUSCH Format 2 (control) is summarized in Table 3. The ACK missed detection probability shall not exceed 1% at the SNR given in table 8.5.2.2.1-2 of [14] for 1Tx case. The ACK missed detection probability is the probability of not detecting an ACK when an ACK was sent per NPUSCH Format 2 transmission. The DTX to ACK probability for NPUSCH Format 2 case denotes the probability that ACK is detected when nothing is sent on the wanted signal and only the noise is present per NPUSCH Format 2 transmission. The minimum requirement for DTX to ACK is defined in section 8.5.2.1.1 of [14] and shall not exceed 1% per NPUSCH Format 2 transmission. The proposed NPUSCH Format 2 receiver meets the performance requirements with a good margin (at least >0.9dB) with the test setup. The results are also in line with the performance reported in [23]. If the implementational complexity is an issue for NPUSCH Format 2, then only the data symbols can be considered in equation (37) for computing the detection metric, with only a minor loss in performance.

| $N_{Tx}$ | $N_{RX}$ | $\Delta f$ | Tones | Channel and correlation matrix | FRC | $N_{rep}$ | Fraction of max. throughput | Minimum SNR at which fraction of throughput should be achieved [14]. | NPUSCH F1 performance with Simulation | NPUSCH F1 performance with Test Setup |
|---|---|---|---|---|---|---|---|---|---|---|
| 1 | 2 | 15KHz | 1 | ETU 1Hz Low | A16-2 | 1 | 70% | -2.1 | -4.5 | -4.2 |
| | | | | | | 16 | 70% | -8.8 | -11.2 | -11 |
| | | | | | | 64 | 70% | -12.6 | -14 | -13.8 |
| | | | 3 | ETU 1Hz Low | A16-3 | 2 | 70% | -3.0 | -7 | -5 |
| | | | | | | 16 | 70% | -8.1 | -12 | -11 |
| | | | | | | 64 | 70% | -11.4 | -15.2 | -14.5 |
| | | | 6 | ETU 1Hz Low | A16-4 | 2 | 70% | -0.6 | -7 | -3 |
| | | | | | | 16 | 70% | -6.8 | -13.5 | -11 |
| | | | | | | 64 | 70% | -10.5 | -16.5 | -14 |
| | | | 12 | ETU 1Hz Low | A16-5 | 2 | 70% | -0.7 | -7 | -3.5 |
| | | | | | | 16 | 70% | -6.4 | -14.5 | -11 |
| | | | | | | 64 | 70% | -10.1 | -17.5 | -15 |

*Table 2: RAN4 Results for NPUSCH Format 1(Data).*

*A. Further enhancements to NB-IoT in Release 15.*

Several important enhancements have been added in Release 15 for NB-IoT (see [25] and [26]) and in this section discussions are provided on the key enhancements finalized in 3GPP Release 15 for NB-IoT. The goal of this section is to show that the enhancements to NB-IoT in Release 15 do not require any significant changes to the receiver design proposed in this paper.

*1) NPRACH reliability and range enhancement:*

The key enhancements made to NPRACH in 3GPP Release 15 are mainly with regards to new preamble formats introduced for range enhancements and time-division multiplexing (TDD) support. For FDD mode, there is one additional preamble format (i.e., Format 2 in frame-structure type 1) defined. For TDD frame-structure type 2, two additional preambles formats (0-a and 1-a) have been defined. A symbol group consists of a cyclic prefix of length $T_{\text{CP}}$ and a sequence of $N$ identical symbols with total length $T_{\text{SEQ}}$. The total number of symbol groups in a preamble repetition unit is denoted by $P$. The number of time-contiguous symbol groups is denoted by $G$. Summary of preamble formats, cyclic prefix and sequence lengths are shown in Table 4 and Table 5 for FDD and TDD respectively. Also, the NPRACH transmission in 3GPP release 15 supports either a 3.75 kHz or 1.25 kHz sub-carrier spacing (only with preamble Format 2 in FDD). The NPRACH formats with different cyclic prefix, symbol group sizes and symbol group repetitions have been added to cater for different ranges and reliability. Symbol groups with $\Delta f_{RA}$ =3.75 kHz (preamble Format 0/1/2 or 0-a/1-a) sub-carrier spacing hop by one or six sub-carriers in frequency as before, and symbol groups with $\Delta f_{RA} = 1.25$ kHz (preamble Format 2 only) subcarrier spacing hop by one, three, or eighteen sub-carriers in frequency. In other words, frequency hopping within one NPRACH preamble is defined for 3-levels of hopping with, -/+ 1.25 kHz, -/+ 3.75 kHz and 22.5 kHz (see [24]), with

- 1.25 kHz hopping gap for 1st to 2nd symbol group and with opposite direction for 5th to 6th symbol group.

| $N_{Tx}$ | $N_{RX}$ | Channel and correlation matrix | Tones | $\Delta f$ | $N_{rep}$ | Minimum SNR in dB at which detection performance is expected [14] | NPUSCH F2 performance with Simulations | NPUSCH F2 performance with Test Setup |
|---|---|---|---|---|---|---|---|---|
| 1 | 2 | EPA 5Hz, Low | 1 | 15KHz | 1 | 6.3 | 3.2 | 4.4 |
| | | | | | 16 | -3.9 | -5.8 | -4.9 |
| | | | | | 64 | -9.5 | -11.8 | -10.6 |

*Table 3: RAN4 Results for NPUSCH Format 2(Control).*

- 3.75 kHz hopping gap for 2nd to 3rd symbol group and with opposite direction for 4th to 5th symbol group.
- 22. 5 kHz hopping gap for 3rd to 4th symbol group.

Repetitions of groups of symbol groups hop by a pseudo-random number of sub-carriers in frequency. There are three in FDD, and four in TDD, possible cyclic prefix lengths for the random-access transmission symbol groups, suitable for different maximum cell sizes (see version 15.3.0 of [10] for more details). The values for NPRACH periodicity are 40, 80, 160, 320, 640, 1280, 2560, 5120 ms. Even with all the above enhancements made to NB-IoT in 3GPP release 15, the NPRACH receiver design proposed in this paper should be agnostic to any changes in the number of symbol group frequency hopping tones. Also, for RTD estimation, in case of $G < P$, the symbol groups may have to be non-coherently combined for RTD estimation if the gap between symbol groups two contiguous symbol groups is larger than the channel coherence time. Note that non-coherent combining may be required for only TDD, since $G = P$ for FDD. With 3-level hopping defined for preamble Format 2, the parameter $T_l$ defined in (14) should be placed appropriately in $F_\rho$. For example, for preamble format 2, modify $T_l$ such that, $\alpha_0 = 1$, $\alpha_1 = 3$, and $\alpha_2 = 18$ and construct the vector $F_\rho$ appropriately for detection.

| Preamble format | $G$ | $P$ | $N$ | $T_{CP}$ | $T_{SEQ}$ |
|---|---|---|---|---|---|
| 0 | 4 | 4 | 5 | 66.67 $\mu s$ or 2048$T_s$ | 5·8192$T_s$ |
| 1 | 4 | 4 | 5 | 266.67 $\mu s$ or 8192$T_s$ | 5·8192$T_s$ |
| 2 | 6 | 6 | 3 | 800 $\mu s$ or 24576$T_s$ | 3·24576$T_s$ |

*Table 4:* NPRACH Preamble formats for FDD in 3GPP Release 15

| Preamble format | $G$ | $P$ | $N$ | $T_{CP}$ | $T_{SEQ}$ |
|---|---|---|---|---|---|
| 0 | 2 | 4 | 1 | 150.5 $\mu s$ or 4778$T_s$ | 1·8192$T_s$ |
| 1 | 2 | 4 | 2 | 266.67 $\mu s$ or 8192$T_s$ | 2·8192$T_s$ |
| 2 | 2 | 4 | 4 | 266.67 $\mu s$ or 8192$T_s$ | 4·8192$T_s$ |
| 0-a | 3 | 6 | 1 | 50 $\mu s$ or 1536$T_s$ | 1·8192$T_s$ |
| 1-a | 3 | 6 | 2 | 100 $\mu s$ or 3072$T_s$ | 2·8192$T_s$ |

*Table 5:* NPRACH preamble formats for TDD in 3GPP Release 15.

*2) NPUSCH Collision with NPRACH:*
In RAN1#95, the following was agreed that for the new NPRACH resources introduced in Release15 with early data transmission (EDT) in uplink.

- For non-EDT NPRACH resources, the UE postpones NPUSCH in those resources only if the UE indicates support for the corresponding feature(s).
- For EDT NPRACH resources, the UE postpones NPUSCH in those resources only during an EDT procedure.

As discussed in section 5.B on the UL gap handling, the block processing for NPUSCH can be stopped and resumed with partial block sizes. Non-coherent combining across partial blocks could be adopted to improve the performance.

*3) Interference randomization for NPUSCH:*
It was identified that NPUSCH channel suffers significant degradation when operating in interference limited scenarios (see [24] and the references therein). Proposals to introduce interference randomization based on SC-FDMA symbol-level scrambling for Rel-14 unicast NPUSCH configured with interference randomization enhancement or Msg3 in non-anchor carrier are currently being discussed. This doesn't impact the NPUSCH receiver design proposed in this paper except that the new data unscrambling module must be introduced after demodulation, whenever it is defined in the 3GPP specification.

## 7. CONCLUSION

In this paper study of the receiver design for NB-IoT uplink channels was presented. Discussions were provided in detail for each uplink channel for time, frequency and signal structures. Receiver design guidelines were provided for all three channels with detailed mathematical analysis. The performance of each channel was investigated and compared against the 3GPP RAN4 requirements with both link level simulations and implementation on a commercially deployed Qualcomm® FSM™ Small Cell platform. Various practical implementation issues were considered and enhancements to NB-IoT in 3GPP Release 15 were discussed. It was shown how the NB-IoT uplink receiver design proposed in this paper can be adapted to include the proposed NB-IoT enhancements in 3GPP Release 15 with minor or no modifications. With NB-IoT being adopted by various network operators across the world, the work in this paper is of significance to system designers looking to implement efficient NB-IoT uplink receiver on practical systems to work in conjunction with legacy LTE processing.

## 8. ACKNOWLEDGEMENTS

The author would like to thank Raja Bachu for his guidance in system design aspects; firmware and test teams in Qualcomm India Private Limited (QIPL), Hyderabad, for all the efforts in developing and testing the proposed receiver algorithms.

## 9. APPENDIX

*A. Definition of mathematical symbols used in derivations.*

| Symbol | Definition |
|---|---|
| $s_{m,i}[n]$ | is the transmit signal at $n^{th}$ sample of $i^{th}$ symbol in $m^{th}$ symbol group |
| $y_{m,i}[n]$ | Received time domain signal of $n^{th}$ sample of $i^{th}$ symbol in $m^{th}$ symbol group |
| $h_{m,i}$ | Channel gain of $i^{th}$ symbol in $m^{th}$ symbol group. |
| $Y_{m,i}[k]$ | Received frequency domain signal of $k^{th}$ tone of $i^{th}$ symbol in $m^{th}$ symbol group |
| $Y_{m,i}$ | Frequency domain samples of $i^{th}$ symbol in $m^{th}$ symbol group. |
| $Y_m$ | Sum of frequency domain samples of $m^{th}$ symbol group. |
| $Z_m$ | Differential metric of $m^{th}$ symbol group. |
| $X_{m1}$ & $X_{m2}$ | Sum and difference of differential metrics across two symbol groups respectively. |
| $W_1$ & $W_2$ | Sum and difference of differential metrics across symbol groups and repetitions respectively. |
| $W$ | Final differential metric with phase proportional to the estimated frequency offset. |
| $V_m$ | Frequency offset corrected differential. |
| $U_{4r+l}$ | Frequency offset corrected and frequency hopping corrected (conjugated) differential. |
| $T_l$ | Frequency offset corrected and frequency hopping corrected (conjugated) differential summed across repetitions. |
| $Q_V$ | FFT vector of corrected differentials. |
| $G_v$ | Raw (not normalized) NPRACH detection metric vector. |
| $G^*$ | Maximum value of the raw (not normalized) NPRACH detection metric vector after polynomial interpolation. |
| $\lambda$ | SNR on the NPRACH subcarrier $\rho$. |
| $N_{rep}^{NPUSCH}$ | Scheduled number of repetitions of a NPUSCH transmission. |
| $N_{symb}^{UL}$ | Number of assigned tones and number of symbols in a slot respectively |
| $N_{slots}^{UL}$ | Number of consecutive slots in an UL resource unit for NB-IoT. |
| $N_{SC}^{RU}$ | Number of consecutive subcarriers in an UL resource unit for NB-IoT |
| $M_{identical}^{NPUSCH}$ | Number of repetitions of identical slots for NPUSCH. |
| $P_k$ | phase difference across adjacent subcarriers $k$ |
| $Q_k^s$ | phase difference across adjacent subcarriers $k$ in slot $s$. |
| $\hat{h}_{s,3}$ | Frequency smoothed channel estimate in slot $s$ for NPUSCH format 1. |
| $H_v$ | FFT vector of frequency smoothed channel estimates. |
| $I_v$ | Raw (not normalized) CFO detection metric. |
| $m_{max}$ | Maximum of CFO detection metric. |
| $\hat{h}_b$ | Average channel estimates for block $b$. |
| $e_{s,l,k}$ | Equalized symbol $l$, for slot $s$, and tone $k$. |
| $e_{s,l,k}^t(n)$ | Time domain sample $n$ of equalized symbol $l$, for slot $s$, and tone $k$. |
| $llr_{s,l,k}$ | Log-likelihood ratios for symbol $l$, for slot $s$, and tone $k$. |
| $SNR_b$ | Signal to noise ratio (SNR) for block $b$. |
| $Z_{s,l}[k_0]$ | Phase de-rotated frequency domain samples for slot $s$ and symbol $l$. |
| $Z_s^p$ | Combined phase de-rotated frequency domain samples for pilot symbols in slot $s$. |
| $Z_s^d$ | Combined phase de-rotated frequency domain samples for data symbols in slot $s$. |
| $J_r(u)$ | Equalized, frequency offset corrected and accumulated pilot and data symbols over RUs and repetitions. |
| $J(u)$ | Detection metric for NPUSCH Format 2 received symbols. |
| $s(t)$ | NB-IoT time domain sample at time $t$. |
| $k$ | NB-IoT tone index. |
| $a_k$ | Modulated alphabet transmitted on tone $k$. |
| $\Delta f$ | Subcarrier spacing between NB-IoT tones. |
| $N_{cp,l}$ | Number of cyclic prefix samples of symbol $l$. |
| $T_s$ | Sampling time interval (Basic Time Unit in [10]). |
| $f_o$ | Frequency separation between centers of LTE and NB-IoT carriers |
| $M_0$ | Time domain samples between beginning of frequency shift application and the $l^{th}$ OFDM symbol. |

## 10. REFERENCES

[1] "Small Cells taking cellular to new heights with IoT tech and global deployments", https://www.qualcomm.com/news/onq/2018/03/16/mwc-2018-small-cells-taking-cellular-new-heights-iot-tech-and-global-deployments , *Mobile World Congress (MWC)*, March 2018.

[2] 3rd Generation Partnership Project, "*NB-LTE – General L1 Concept Description*", Ericsson, et.al, R1-156010, Oct. 2015.

[3] Y.-P.Eric Wang et.al, " *A Primer on 3GPP Narrowband Internet of Things (NB-IoT)"*, IEEE Communication magazine Vol 55, issue 3, March 2017.

[4] R. Ratasuk, et.al*, "Overview of narrowband IoT in LTE Rel-13,"* IEEE Conference on Standards for Communications and Networking (CSCN), Berlin, 2016, pp. 1-7.

[5] X. Lin, A. Adhikary, and Y.-P. E. Wang*, "Random Access Preamble Design and Detection for 3GPP Narrowband IoT systems,"* IEEE Wireless Communications Letters, vol. 5, no. 6, pp. 640–643, Dec. 2016.

[6] R. Harwahyu, R. G. Cheng, C. H. Wei and R. F. Sari, "*Optimization of Random Access Channel in NB-IoT*," in IEEE Internet of Things Journal, vol. 5, no. 1, pp. 391-402, Feb. 2018.

[7] R. Ratasuk, et.al*, "Data Channel Design and Performance for LTE Narrowband IoT,"* IEEE 84th Vehicular Technology Conference (VTC-Fall), Montreal, QC, 2016, pp. 1-5.

[8] J. Hwang, C. Li and C. Ma*, "Efficient Detection and Synchronization of Superimposed NB-IoT NPRACH Preambles,"* in IEEE Internet of Things Journal. vol. 6, no. 1, pp. 1173-1182, Feb. 2019.

[9] 3rd Generation Partnership Project, *"NB-PRACH design- document for discussion and decision"*, R1-160316, Huawei, HiSilicon, Neul, 3GPP, TSG RAN WG1 Meeting #84, Feb 2016.

[10] 3rd Generation Partnership Project, "*Technical Specification Group Radio Access Network; Evolved Universal Terrestrial Radio Access; Physical Channels and Modulation*", *3GPP TS 36.211 V14.1.0.*, Dec. 2016.

[11] 3rd Generation Partnership Project, "*Technical Specification Group Radio Access Network; Evolved Universal Terrestrial Radio Access; Physical layer Procedures*", 3GPP TS 36.213 V14.1.0., Dec. 2016.

[12] 3rd Generation Partnership Project, "*Technical Specification Group Radio Access Network; Evolved Universal Terrestrial Radio Access; Radio Resource Control (RRC), Protocol specification*", 3GPP TS 36.331 V14.2.0., Mar. 2017.

[13] 3rd Generation Partnership Project, "*Technical Specification Group Radio Access Network; Evolved Universal Terrestrial Radio Access; Medium Access Control (MAC) protocol specification*", 3GPP TS 36.321 V14.1.0., Mar. 2016.

[14] 3rd Generation Partnership Project, "*Technical Specification Group Radio Access Network; Evolved Universal Terrestrial Radio Access; Base Station (BS) radio transmission and reception*", 3GPP TS 36.104 V14.5.0., Sep. 2017.

[15] 3rd Generation Partnership Project, "*Technical Specification Group Radio Access Network; Evolved Universal Terrestrial Radio Access; Base Station (BS) conformance testing*", *3GPP TS 36.141 V14.5.0.*, Sep. 2017.

[16] J.O. Smith, "Spectral Audio Signal Processing", W3K Publishing, http://books.w3k.org/, ISBN 978-0-9745607-3-1.

[17] 3rd Generation Partnership Project, "*Technical Specification Group Radio Access Network; Evolved Universal Terrestrial Radio Access; Radio Resource Control (RRC): Protocol Specification*", *3GPP TS 36.331 V14.2.0.*, March. 2017.

[18] R. Ratasuk, N. Mangalvedhe, Z. Xiong, M. Robert and D. Bhatoolaul, "Enhancements of narrowband IoT in 3GPP Rel-14 and Rel-15," IEEE Conference on Standards for Communications and Networking (CSCN), Helsinki, 2017, pp. 60-65.

[19] A. Chakrapani, "*Efficient resource scheduling for eMTC/NB-IoT communications in LTE Rel. 13*," IEEE Conference on Standards for Communications and Networking (CSCN), Helsinki, 2017, pp. 66-71.

[20] Keysight Signal generator , MXG N5182B, web link https://literature.cdn.keysight.com/litweb/pdf/5990-9959EN.pdf.

[21] Vertex Channel emulator, VR5, web link, https://www.spirent.com/-/media/Datasheets/Mobile/VR5_Datasheet.pdf

[22] 3rd Generation Partnership Project, *'Summary of simulation results for NB-IoT BS demodulation requirement',* R4-166484, 3GPP TSG-RAN WG4 meeting #80, Aug. 2016.

[23] 3rd Generation Partnership Project," *Feature lead summary NPRACH reliability and range enhancements"*, R1-1807429, 3GPP TSG-RAN WG1 Meeting #93, May 2018.

[24] R. Ratasuk, et.al, "Enhancements of narrowband IoT in 3GPP Rel-14 and Rel-15", IEEE Conference on Standards for Communications and Networking (CSCN), pp. 60-65, Sept 2017.

[25] 3rd Generation Partnership Project, "*Technical Specification Group Radio Access Network; Evolved Universal Terrestrial Radio Access; and Evolved Universal Terrestrial Radio Access Network, Overall Description*", 3GPP TS 36.331 V14.2.0., Mar. 2017.